\documentclass[12pt,draftclsnofoot,onecolumn,peerreview,journal]{IEEEtran}
\usepackage{graphicx}
\usepackage{verbatim,cite,flafter,caption,amsmath,amssymb}
\usepackage{mathrsfs}
\makeatletter
\def\fps@figure{hbp}
\makeatother
\begin{document}
\title{Generalized reliability-based syndrome decoding for LDPC codes}
\author{Guangwen Li, Guangzeng Feng}
\maketitle
\bibliographystyle{ieeetran}
\begin{abstract}
Aiming at bridging the gap between the maximum likelihood decoding
(MLD) and the suboptimal iterative decodings for short or medium
length LDPC codes,  we present a generalized ordered statistic
decoding (OSD) in the form of syndrome decoding, to cascade with the
belief propagation (BP) or enhanced min-sum decoding. The OSD is
invoked only when the decoding failures are obtained for the
preceded iterative decoding method. With respect to the existing OSD
which is based on the accumulated log-likelihood ratio (LLR) metric,
we extend the accumulative metric to the situation where the BP
decoding is in the probability domain. Moreover, after generalizing
the accumulative metric to the context of the normalized or offset
min-sum decoding, the OSD shows appealing tradeoff between
performance and complexity. In the OSD implementation, when deciding
the true error pattern among many candidates, an alternative
proposed proves to be effective to reduce the number of real
additions without performance loss. Simulation results demonstrate
that the cascade connection of enhanced min-sum and OSD decodings
outperforms the BP alone significantly, in terms of either
performance or complexity.
\end{abstract}
\IEEEpeerreviewmaketitle
\section{Introduction}
For finite length low-density parity-check (LDPC) codes, the
effective belief propagation (BP) decoding is regarded as
suboptimal, owing to the presence of unavoidable short cycles in the
bipartite graph representation of the code. On the other hand, due
to the exponentially increased complexity with the block length,
there exists no feasible maximum likelihood decoding (MLD) for
practical LDPC codes.

To bridge the gap between BP and MLD decodings for short or medium
length LDPC codes, a reliability-based order statistic decoding
(OSD) was proposed to combine with the BP decoding
\cite{fossorier2001irb_2} \cite{fossorier2001irb}. For such a BP-OSD
reprocessing strategy, if no valid codeword is found at some
iteration of the BP decoding, the delivered reliability information
is used as the input to the OSD. Then the OSD decides whether one
more iteration of the BP decoding is necessary according to some
rule. For most codes of interest, the simulations have shown that BP
decoding, in conjunction with order-$p$ OSD reprocessing, yields
near-optimal decoding performance when $p=4$
\cite{fossorier2001irb}. Unfortunately,  the incurred complexity in
each OSD invoking  boosts rapidly with the block length, hence
greatly exceeds the complexity of  one iteration of BP decoding
\cite{isaka2002sid}\cite{isaka2004sid}. On the other hand, after
approximating complex $\tanh$ function with simple $\min$ function
for the check node updating, the min-sum algorithm
\cite{fossorier1999rci} reduces the BP decoding complexity
substantially. Furthermore, its enhanced variants, the normalized or
offset min-sum algorithm \cite{chen2005rcd}, achieves comparable
performance as the BP decoding.

Despite these efforts, there is still space to improve with respect
to the tradeoff between performance and complexity. In
\cite{fossorier2001irb}, it was conjectured that, owing to the
inaccurate reliability information delivered at the last iteration
of BP decoding, order-$p$ OSD in cascade connection with the BP
would result in negligible improvement over the BP alone, where the
word ''cascade'' implies one invoking of OSD per sequence in the
paper. Nevertheless, it was shown in \cite{jiangm2007} that such
cascade connection could achieve noticeable performance gain by
drawing upon the accumulative log-likelihood ratio (LLR)
information, instead of using only the LLR information of the last
iteration of BP decoding as mentioned in \cite{fossorier2001irb}.

When the BP is implemented in probability domain, the available
reliability information about each codeword bit is the probability
of it being one or zero, we thus devise a similar accumulative
probability metric, without converting it to the equivalent LLR
representation, and hoping that it could achieve comparable
performance as \cite{jiangm2007}. When the idea of accumulated
reliability metric is applied in the context of reduced complexity
decoding, say normalized or offset min-sum decoding, with respect to
the BP alone, such a min-sum plus OSD decoding is justified by
significant performance improvement and much less complexity.

The rest of the paper is organized as follows. Section \makeatletter
\@Roman{2} \makeatother explains the steps of OSD based on various
reliability metrics. Section \makeatletter \@Roman{3} \makeatother
details the simulation results and discussion. Finally we conclude
the work in Section \makeatletter \@Roman{4} \makeatother.

\section{Implementation of the syndrome decoding}
Assume a high rate binary $(N,K)$ LDPC code with length $N$ and
dimension $K$. The parity check matrix is of the form
$H_{M\times{N}}$, where $M=N-K$. The
BPSK modulation maps one codeword $c=[c_1,c_2,\ldots,c_N]$ into
$x=[x_1,x_2,\ldots,x_N]$ with $x_i=2c_i-1$, where $i=1,2,\ldots,N$.
After the symbols are transmitted through an additive white Gaussian
noise (AWGN) memoryless channel, we obtain the corrupted sequence
$y=[y_1,y_2,\ldots,y_N]$ at the receiver, where $y_i=x_i+z_i$, $z_i$
is an independent Gaussian random variable with a distribution of
$\mathscr{N}(0,\sigma^2)$.

Given the bipartite graph of a code, the BP or enhanced min-sum
decoding iteratively exchanges message between variable nodes and
check nodes, the tentative hard decision is made after each
iteration to test whether all check sums are satisfied. If so, exit
immediately and declare decoding success. Otherwise continue
the iterative decoding till the maximum number of iteration $I_m$ is
reached. The interested readers could refer to \cite{chen2005rcd} \cite{richardson2001}
 for more detailed description.

For the BP-OSD reprocessing structure mentioned in
\cite{fossorier2001irb}, to achieve near optimal performance,
multiple invokings of OSD are required per sequence, which may
result in unbearable decoding delay due to the high complexity
attributed to the order-$p$ OSD.  While in the BP-OSD or min-sum-OSD
cascade connection addressed in \cite{jiangm2007} and the paper, one
invoking of OSD is needed only when  no valid codeword is returned
after the $I_m$th iterative decoding.

While in \cite{fossorier2001irb} \cite{fossorier1995sdd} , the OSD is
involved with the most reliable basis (MRB) of generator matrix $G$.
In \cite{fossorier1998rbs}, it has been proved that the least
reliable basis (LRB) of $H$ and MRB of $G$ are
dual of each other. Thus, the OSD, in the form of syndrome decoding,
has equivalent error performance as that in
\cite{fossorier1995sdd}. After taking into account the lower
dimension and sparseness of $H$, we prefer the OSD in the form of
syndrome decoding, because it is easier to secure the LRB of $H$ than the MRB of $G$.

When the BP is implemented in LLR domain, the accumulative
reliability metric for variable node $i$ is defined as
\cite{jiangm2007}
\begin{equation}
\label{jiang_def}
 r_i=\sum_{k=0}^{I_m}\alpha^{I_m-k}L_i^{(k)}
\end{equation}
where $\alpha$ is a weight factor the optimal value of which may
resort to the simulation. $L_i^{(k)}$ is the LLR output of variable
node $i$ at the $k$th iterative decoding, with
$$
L_i^{(0)}=\ln
\frac{P'(y_i|c_i=1)}{P'(y_i|c_i=0)}=\frac{2y_i}{\sigma^2}
$$
where $P'(\cdot)$ is the conditional probability. Hence the
hard-decision on variable node $i$ is
$$
\hat{c}_i=\begin{cases} 1 & \text{if } r_i>0,\\
0 & \text{if } r_i\leq0.
\end{cases}
$$

Similarly, when the BP is in probability domain, assume the
syndrome of the initial hard decision on $y$ is $\hat{c}$, after
each iteration of BP decoding, the accumulative probability metric
for variable node $i$ is defined as
\begin{equation}
\label{li_def}
 q_i=\begin{cases} \sum_{k=0}^{k=I_m}\alpha^{I_m-k}P_i^{(k)}&\text{if } \hat{c}_i=1,\\
 I_m+1-\sum_{k=0}^{k=I_m}\alpha^{I_m-k}P_i^{(k)}&\text{if } \hat{c}_i=0
\end{cases}
\end{equation}
where $P_i^{(k)}$ is the probability of variable $i$ being one at
the $k$th iterative decoding.

Furthermore, when the normalized or offset min-sum decoding is
substituted for the BP decoding, we extend the idea of accumulative
reliability metric for the corresponding OSD as well.
\begin{equation}
\label{min_sum_def}
 u_i=\sum_{k=0}^{I_m}\alpha^{I_m-k}U_i^{(k)}
\end{equation}
where $U_i^{(k)}$ is the reliability  measurement of variable node
$i$ at the $k$th iterative decoding, with $y_i$ being the initial
$U_i^{(0)}$. The hard-decision on $u_i$ for variable node $i$ is defined as
$$
\hat{c}_i=\begin{cases} 1 & \text{if } u_i>0,\\
0 & \text{if } u_i\leq0.
\end{cases}
$$
Since the enhanced min-sum variants are known as
uniformly most powerful (UMP) decoding, in the sense that no noise
characteristic about channel is required, the same advantage is
manifested for the OSD when drawing upon the definition of (\ref{min_sum_def}).

For the OSD, its information set includes all bits except those in
the LRB. And it is possible for order-$p$ OSD to solve the
decoding failures when at most $p$ erroneous bits are included in
the information set. Hence, the construction of LRB and information
set, dependent on the reliability evaluation of each codeword bit,
determines the OSD performance. Given a specific code and a small number $p$,
one reliability metric is said to be preferred to another in the sense
that more decoding failures are within the scope of
order-$p$ OSD, considering that the estimated error pattern $\hat{e}$ which  satisfies the following discrepancy test
\cite{fossorier2001irb} has a high probability to be the true error pattern.
\begin{equation}
\label{judge_criterion}
D(y,\hat{e})=\min{\sum_{i:\hat{e}_i=1}|y_i|}
\end{equation}

Based on the decoding framework presented in \cite{fossorier1995sdd,jiangm2007,fossorier1998rbs,fossorier2001irb},
with a modification only at step 4 below, the BP or enhanced min-sum variant, in
cascade connection with the OSD, proceeds as follows.
\begin{enumerate}
\item
The OSD  is invoked iff  all check sums are not satisfied
after the $I_m$th iteration of BP or enhanced min-sum decoding.
\item
With the reliability evaluation obtained with (\ref{jiang_def}),
(\ref{li_def}) or (\ref{min_sum_def}), dependent on the iterative
decoding method. Permutation $\lambda_1$ sorts  the bits $\hat{e}_i,
i \in [1,N]$ of error pattern $\hat{e}$ in increasing order of the
absolute reliability value, hence $H$ is transformed into $H^{(1)}$
by column reordering. Permutation $\lambda_2$ on $H^{(1)}$ is to
ensure the leftmost $M$ columns of resultant $H^{(2)}$ to be
independent, the indices of which form the LRB, and the other bit
positions make up the information set. Accordingly,
$\hat{e}^{(2)}=\lambda_2(\lambda_1(\hat{e}))$.
\item
Apply Gaussian elimination on $H^{(2)}$ to transform it into
systematic form. That is
\begin{equation}
\label{equ:matrix_represent}
EH^{(2)}\hat{e}^{(2)}=E\hat{c}\Rightarrow[I \hspace{3pt} EH^{(2)}_{12}][\hat{e}^{(2)}_{11} \hspace{3pt} \hat{e}^{(2)}_{12}]'
=E\hat{c}\Rightarrow\hat{e}^{(2)}_{11}=EH^{(2)}_{12}\hat{e}^{(2)}_{12}+E\hat{c}
\end{equation}
where $E$ is the equivalent matrix for Gaussian elimination operation.
\item
Evidently, $\hat{e}=\lambda_1^{-1}(\lambda_2^{-1}(\hat{e}^{(2)})$. For order-$p$ OSD, totally $\sum_{i=0}^{i=p}\binom{K}{i}$ candidate error patterns
were supposed to be traversed. Conventionally, (\ref{judge_criterion}) will be solicited
for each candidate to seek the one with the minimum discrepancy. To lower the complexity, we divide the task of seeking the optimal error pattern
into two substeps.
\begin{enumerate}
\item
Assign a weight $w_i$ for each bit $i$ while transforming $H$ into $H^{(1)}$ in step 2. Specifically, if bit $i$ is in the $j$th position after reordering,
it will be evaluated $w_i=j$. The use the metric $W_s=\sum_{i:\hat{e}_i=1}w_i$ to find $\beta$ candidates which have the
smallest $W_s$.
\item
Apply (\ref{judge_criterion}) to make a decision  among the $\beta$ error patterns. When a code has appropriate minimum distance,
 no performance loss is observed in the simulation even if $\beta=1$.
\end{enumerate}
In such a way, many real additions in computing (\ref{judge_criterion}) otherwise are replaced by simple integer additions.
\item
Apply modulo $2$ addition of $\hat{e}$ with $\hat{c}$ to recover the original codeword.
\end{enumerate}

\section{Simulation results and Discussion}
The simulations are performed on a number of LDPC codes, but we will
present only the results for $(504,252)$ \cite{mackay_database}. At
each SNR point, at least $100$ decoding failures are detected. To
manage computational complexity of the OSD, the simulations are
limited to at most order-2 OSD. (\ref{jiang_def}), (\ref{li_def}) and (\ref{min_sum_def}) all relate to
the evaluation of parameter $\alpha$. As reported in \cite{jiangm2007}, the optimal value of $\alpha$ relies on
$I_m$ and the code itself. For $(504,252)$, it was found $\alpha=1$ results in the best decoding performance when $I_m=20$.
Therefore, For the sake of simplicity, $\alpha=1$ is assumed in the simulations below.

\begin{figure}[h]
\centering
\includegraphics[width=0.75\linewidth]{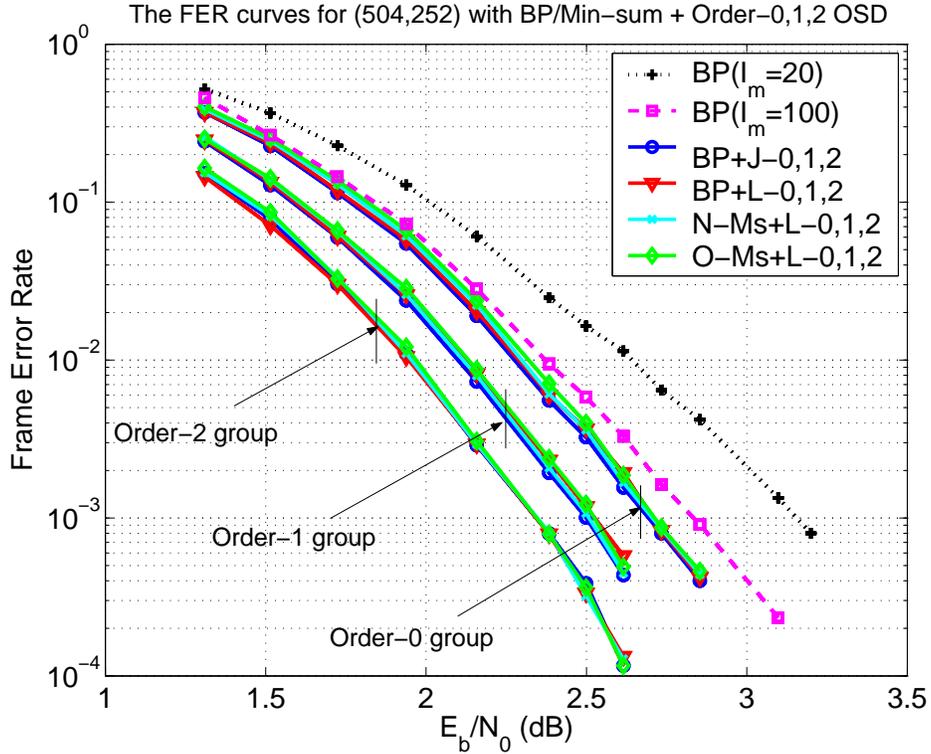}
\parbox{0.95\linewidth}{\caption{\label{fig:504_decoding_FER} The FER curves for (504,252) with
BP/Min-sum + Order-0,1,2 OSD decoding schemes of various reliability
metrics}}
\end{figure}

The frame error rate (FER) curves for each iterative decoding, all
with $I_m=20$, in cascade connection separately with order-0, 1, 2 OSD, are
depicted in \ref{fig:504_decoding_FER}, also included are FER curves
for the BP decoding with $I_m=20, 100$, respectively. In the legend,
BP+J-$0, 1, 2$ denotes the BP cascaded separately with
order-$0, 1, 2$ OSD, whose reliability metric is given by
(\ref{jiang_def}). While for the OSD in BP+L-$0, 1, 2,$ its reliability
metric is given by (\ref{li_def}).  (\ref{min_sum_def}) is used in N-Ms+L-$0, 1, 2$ and O-Ms+L-$0, 1, 2,$ where
''N-Ms'' stands for the normalized min-sum, and ''O-Ms'' the offset min-sum.

\begin{figure}[h]
\begin{minipage}[t]{0.51\linewidth}
\centering
\includegraphics[width=0.95\linewidth]{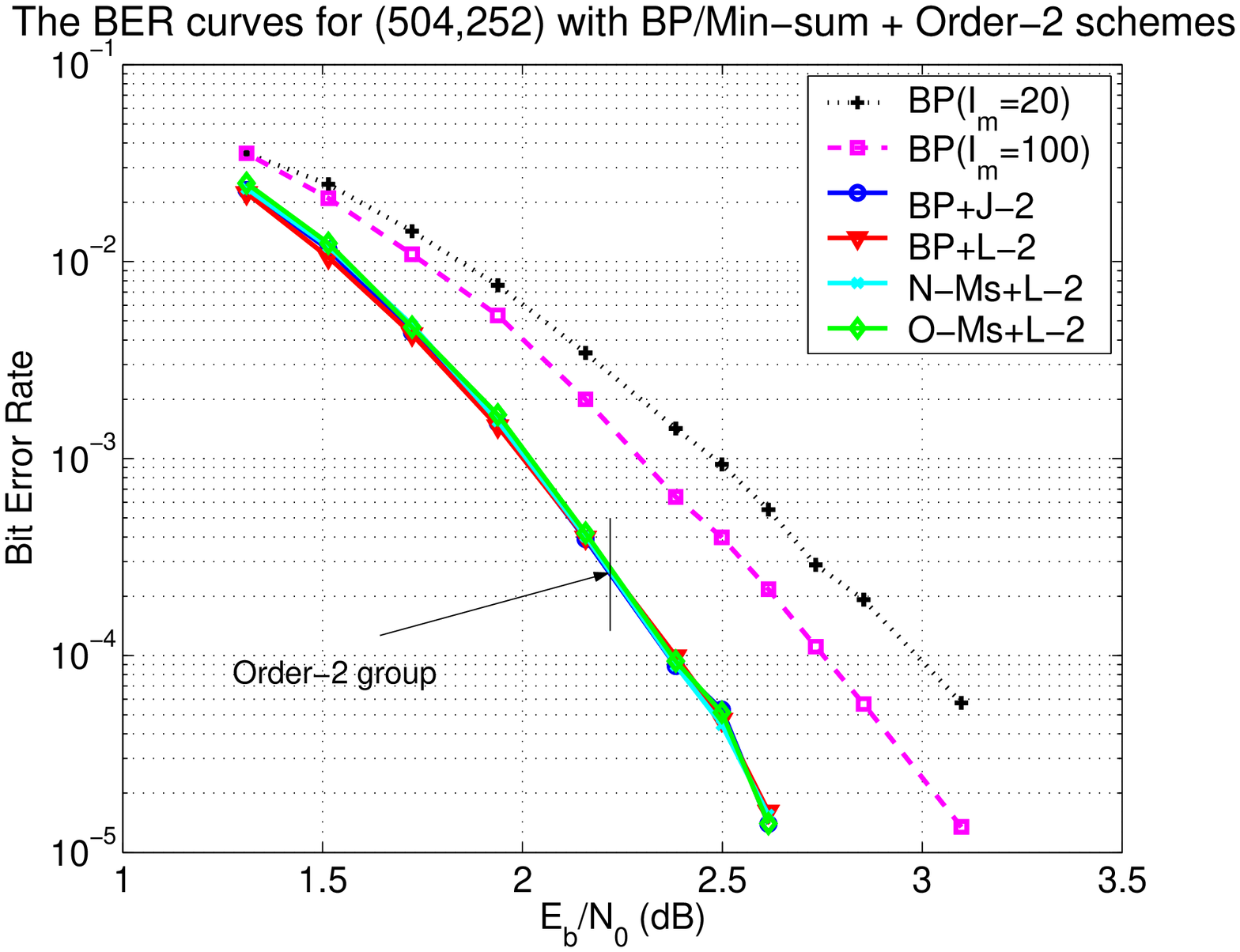}
\parbox{0.95\linewidth}{\caption{\label{fig:504_decoding_BER} The FER curves for (504,252) with
BP/Min-sum + Order-2 OSD decoding schemes of various reliability
metrics}}
\end{minipage}%
\begin{minipage}[t]{0.49\linewidth} \centering
\includegraphics[width=0.95\linewidth]{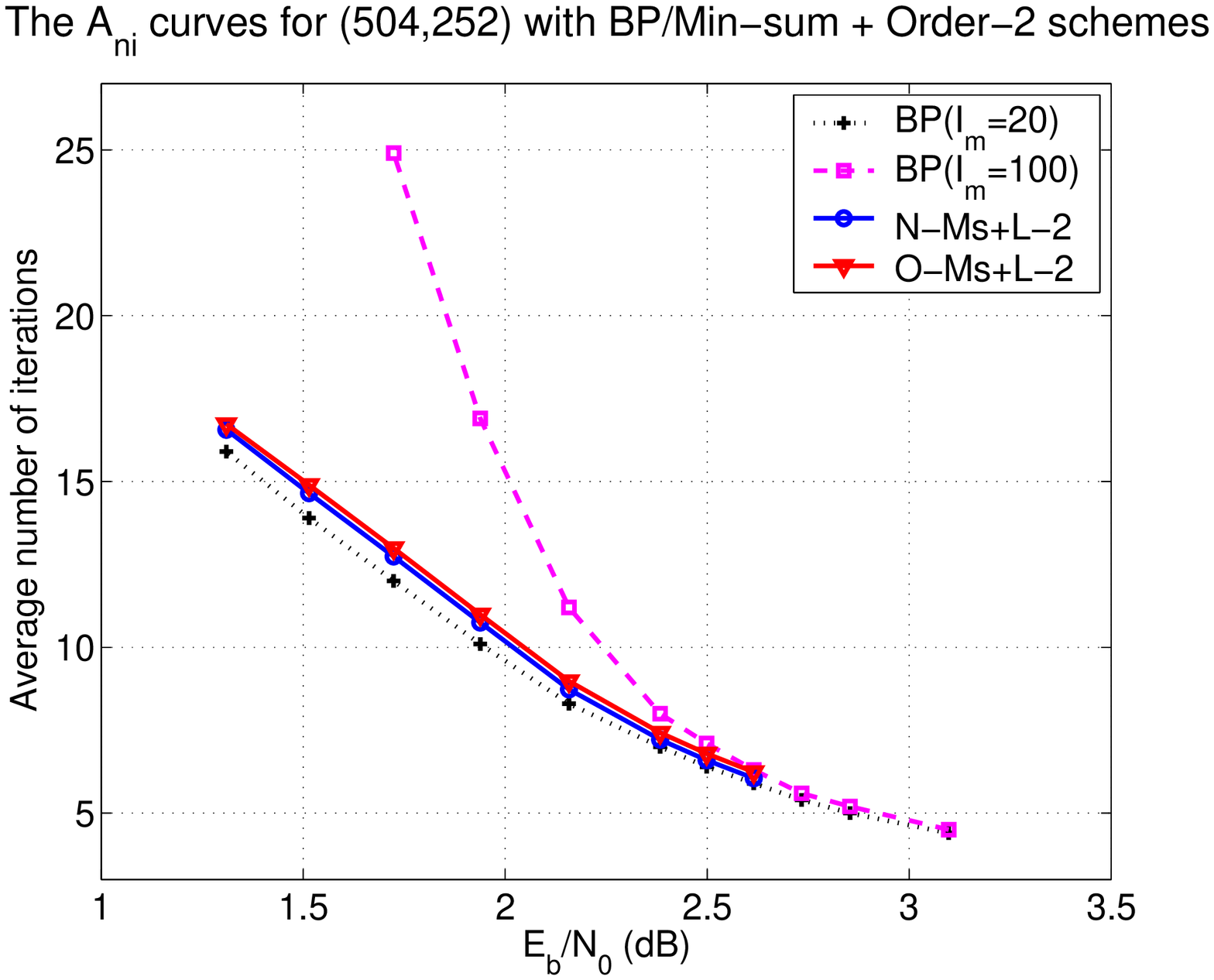}
\parbox{0.95\linewidth}{\caption{\label{fig:504_decoding_iterations} The $A_{ni}$ curves for (504,252) with
BP/Min-sum + Order-2 OSD decoding schemes of various reliability
metrics}}
\end{minipage}
\end{figure}

As seen in \ref{fig:504_decoding_FER} and \ref{fig:504_decoding_BER}, Whatever in the FER or bit error rate (BER) metric,
given a specific $p$, there exists no performance difference among each BP/Min-sum plus
order-$p$ OSD combination. Specifically, in \ref{fig:504_decoding_FER}, the BP+J-0, BP+L-0,
N-Ms+L-0 and O-Ms+L-0, indicated as Order-0 group, all achieve $0.5$\,dB at FER=$10^{-3}$ over the BP ($I_m=20$)
and $0.12$\,dB over the BP ($I_m=100$). While the Order-$1, 2$ groups outperform the BP ($I_m=100$) by $0.35, 0.50$\,dB, respectively.

From the perspective of BER, as shown in \ref{fig:504_decoding_BER}, at BER=$10^{-4}$, Order-2 group
achieves $0.65, 0.45$\,dB over the BP with $I_m=20,100$, respectively. While Order-$0, 1$ groups
achieve observable performance improvement over the BP alone as well, the BER curves of which are not depicted
in \ref{fig:504_decoding_BER}.

The OSD always returns an valid codeword, nevertheless this codeword may contain more erroneous bits
than the decoding failure when it is not the correct one.
Therefore, the performance gain achieved in BER metric is not so striking as in FER metric assuming the
 same decoding combination, as reported in \cite{isaka2004sid} as well.

Denote the average number of iterations as $A_{ni}$, it is shown in \ref{fig:504_decoding_iterations} that in the SNR region of
interest, the $A_{ni}$ required for each decoding combination is less than that of the BP ($I_m=100$) alone. Taking into account \ref{fig:504_decoding_FER}, \ref{fig:504_decoding_BER}
and \ref{fig:504_decoding_iterations}, we find that the normalized or offset min-sum, as a reduced complexity decoding, when in cascade connection with the order-$p$ OSD,
appears to be the most competitive decoding scheme, in terms of tradeoff between  performance and complexity.

For the detailed complexity analysis about the BP in probability and LLR domain, and normalized or
offset min-sum, the interested readers could refer to
\cite{fossorier2001irb} \cite{chen2005rcd}. Among these four schemes, the offset min-sum
requires the least complexity. To obtain the reliability evaluation for all codeword bits,
$NI_m$ real additions are required during iterative decoding to accumulate it.
As far as the proposed OSD is concerned, the binary operations for Gaussian elimination is of the
order $O(N^3)$, the integer and binary operations for each phase-$l$ ($1\leq l\leq p$) of order-$p$ OSD is of
the order $O(N^{l+1})$. Also included are $N\log_2N$ real additions for sorting codeword bits and
$\beta\gamma$ real additions for estimating the error pattern, where $\gamma$ is the average number of nonzero
element in an error pattern.

\section{Conclusions}
In this paper, we have generalized the BP-OSD postprocessing
framework into more applications. For the order-$p$ OSD, we extend
the accumulated reliability metric which was applied in the LLR
domain of BP decoding, to the probability domain of BP decoding.
Furthermore, an extension of accumulated metric to the reduced
complexity decoding, say min-sum variants, has shown that such
combination will be the most advantageous, in the sense that no
channel characteristic is required and the best tradeoff is achieved
between performance and complexity.

Given a fixed $p$, the performance of an order-$p$ OSD will suffer
from the block length increment. How to mitigate such degradation
remains to be solved in the future work.

\section*{Acknowledgment}
Sincere thanks to R.Neal for his online
software\cite{neal_software}. This work was supported by National
Science Foundation of China under Grant No.60472104.

\bibliography{IEEEabrv,output1}

\begin{thebibliography}{10}
\providecommand{\url}[1]{#1}
\csname url@rmstyle\endcsname
\providecommand{\newblock}{\relax}
\providecommand{\bibinfo}[2]{#2}
\providecommand\BIBentrySTDinterwordspacing{\spaceskip=0pt\relax}
\providecommand\BIBentryALTinterwordstretchfactor{4}
\providecommand\BIBentryALTinterwordspacing{\spaceskip=\fontdimen2\font plus
\BIBentryALTinterwordstretchfactor\fontdimen3\font minus
  \fontdimen4\font\relax}
\providecommand\BIBforeignlanguage[2]{{%
\expandafter\ifx\csname l@#1\endcsname\relax
\typeout{** WARNING: IEEEtran.bst: No hyphenation pattern has been}%
\typeout{** loaded for the language `#1'. Using the pattern for}%
\typeout{** the default language instead.}%
\else
\language=\csname l@#1\endcsname
\fi
#2}}

\bibitem{fossorier2001irb_2}
M.~Fossorier, ``{Iterative reliability based decoding of LDPC codes},''
  \emph{Information Theory, 2001. Proceedings. 2001 IEEE International
  Symposium on}, 2001.

\bibitem{fossorier2001irb}
------, ``{Iterative reliability-based decoding of low-density parity
  checkcodes},'' \emph{Selected Areas in Communications, IEEE Journal on},
  vol.~19, no.~5, pp. 908--917, 2001.

\bibitem{isaka2002sid}
M.~Isaka, M.~Fossorier, and H.~Imai, ``{On the suboptimality of iterative
  decoding for finite length codes},'' \emph{Information Theory, 2002.
  Proceedings. 2002 IEEE International Symposium on}, 2002.

\bibitem{isaka2004sid}
------, ``{On the suboptimality of iterative decoding for turbo-like and LDPC
  codes with cycles in their graph representation},'' \emph{Communications,
  IEEE Transactions on}, vol.~52, no.~5, pp. 845--854, 2004.

\bibitem{fossorier1999rci}
M.~Fossorier, M.~Mihaljevic, and H.~Imai, ``{Reduced complexity iterative
  decoding of low-density parity checkcodes based on belief propagation},''
  \emph{Communications, IEEE Transactions on}, vol.~47, no.~5, pp. 673--680,
  1999.

\bibitem{chen2005rcd}
J.~Chen, A.~Dholakia, E.~Eleftheriou, M.~Fossorier, and X.~Hu,
  ``{Reduced-Complexity Decoding of LDPC Codes},'' \emph{Communications, IEEE
  Transactions on}, vol.~53, no.~8, pp. 1288--1299, 2005.

\bibitem{jiangm2007}
M.~Jiang, C.~Zhao, E.~Xu, and L.~Zhang, ``{Reliability-Based Iterative Decoding
  of LDPC Codes Using Likelihood Accumulation},'' \emph{Communications Letters,
  IEEE}, vol.~11, no.~8, pp. 677--679, 2007.

\bibitem{richardson2001}
R.~Richardson, T.~Urbanke, ``The capacity of low-density parity-check codes
  under message-passing decoding,'' \emph{{IEEE} Transactions on Information
  Theory}, vol.~47, pp. 599--618, 2001.

\bibitem{fossorier1995sdd}
M.~Fossorier and S.~Lin, ``{Soft-decision decoding of linear block codes based
  on ordered statistics},'' \emph{Information Theory, IEEE Transactions on},
  vol.~41, no.~5, pp. 1379--1396, 1995.

\bibitem{fossorier1998rbs}
M.~Fossorier, S.~Lin, and J.~Snyders, ``{Reliability-based syndrome decoding of
  linear block codes},'' \emph{Information Theory, IEEE Transactions on},
  vol.~44, no.~1, pp. 388--398, 1998.

\bibitem{mackay_database}
\BIBentryALTinterwordspacing
D.~J.~C. MacKay, ``{Encyclopedia of Sparse Graph Codes}.'' [Online]. Available:
  \url{http://www.inference.phy.cam.ac.uk/mackay/CodesFiles.html}
\BIBentrySTDinterwordspacing

\bibitem{neal_software}
\BIBentryALTinterwordspacing
R.~M. Neal, ``{Software for low-density parity-check codes}.'' [Online].
  Available: \url{http://www.cs.toronto.edu/~radford/ldpc.software.html}
\BIBentrySTDinterwordspacing

\end{thebibliography}
\end{document}